# Visualization of Mined Pattern and Its Human Aspects

Ratnesh Kumar Jain, Dr. R. S. Kasana
Department of Computer Science and Applications
Dr. H. S. Gour Central University
Sagar, MP, India.
jratnesh@rediffmail.com, irkasana7158.gmail.com

Dr. Suresh Jain
Department of Computer Engineering, Institute of Engineering & Technology,
Devi Ahilya University, Indore, MP (India)
suresh.jain@rediffmail.com

*Abstract*—Researchers got success in mining the Web usage data effectively and efficiently. But representation of the mined patterns is often not in a form suitable for direct human consumption. Hence mechanisms and tools that can represent mined patterns in easily understandable format are utilized. Different techniques are used for pattern analysis, one of them is visualization. Visualization can provide valuable assistance for data analysis and decision making tasks. In the data visualization process, technical representations of web pages are replaced by user attractive text interpretations. Experiments with the real world problems showed that the visualization can significantly increase the quality and usefulness of web log mining results. However, how decision makers perceive and interact with a visual representation can strongly influence their understanding of the data as well as the usefulness of the visual presentation. Human factors therefore contribute significantly to the visualization process and should play an important role in the design and evaluation of visualization tools. This electronic document is a "live" template. The various components of your paper [title, text, heads, etc.] are already defined on the style sheet, as illustrated by the portions given in this document.

## I. INTRODUCTION

The dictionary meaning of visualize is "to form a mental vision, image, or picture of (something not visible or present to sight, or of an abstraction); to make visible to the mind or imagination" [The Oxford English Dictionary, 1989]. The discovery of Web usage patterns would not be very useful unless there are mechanisms and tools to help an analyst better understand them. Visualization has been used very successfully in helping people understand various kinds of phenomena both real and abstract. Hence it is a natural choice for understanding the behavior of Web users. "The essence of Information Visualization is referred to the creation of an internal model or image in the mind of a user. Hence, information visualization is an activity that humankind is engaged in all the time". [1]

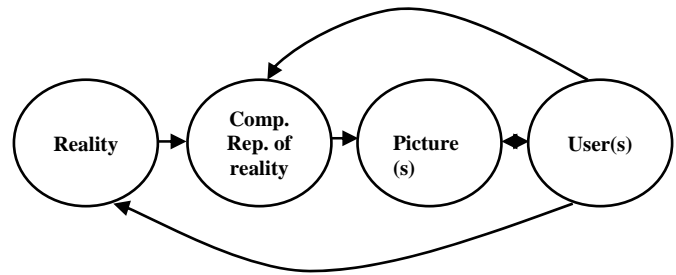

Figure 1. Visualization Process

Visualization of the web usage data is a technique in which mined data are represented graphically. In this process, technical representations of web pages are replaced by user attractive text interpretations.

### A. VISUALIZATION TECHNIQUES

There are a large number of visualization techniques which can be used for visualizing the data. In addition to standard 2D/3D-techniques, such as x-y (x-y-z) plots, bar charts, line graphs, etc., there are a number of more sophisticated visualization techniques (see fig. 2). The classes correspond to basic visualization principles which may be combined in order to implement a specific visualization system.

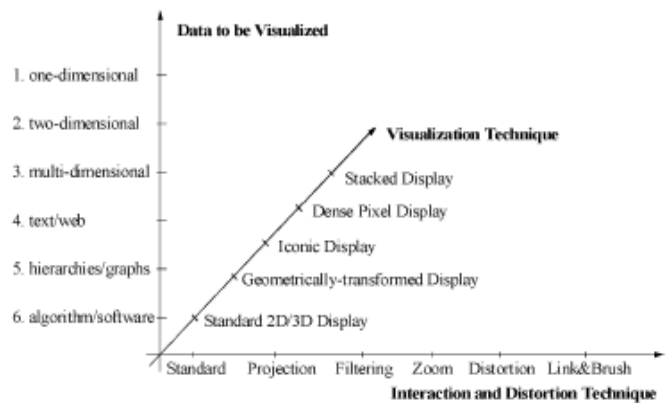

Figure 2. Classification of Visualization technique

*1) Geometrically Transformed Displays*

Geometrically transformed display techniques aim at finding "interesting" transformations of multidimensional data sets. The class of geometric display techniques includes techniques from exploratory statistics, such as scatter plot matrices and techniques which can be subsumed under the term "projection pursuit". Other geometric projection techniques include Projection Views, Hyperslice, and the well-known Parallel Coordinates visualization technique.

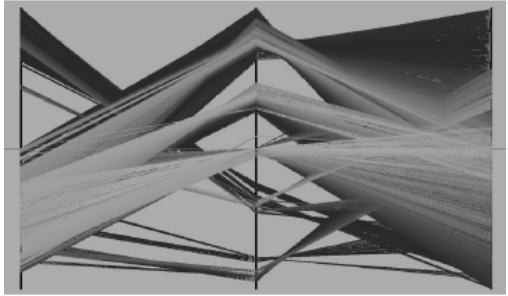

Figure 3. Parallel Coordinate Visualization

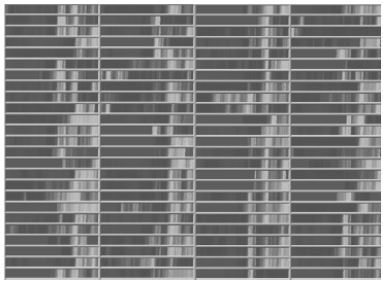

Figure 4. Dense pixel displays (Courtesy IEEE)

*2) Iconic Displays*

Another class of visual data exploration techniques is the iconic display techniques. The idea is to map the attribute values of a multidimensional data item to the features of an icon.

*3) Dense Pixel Displays*

The basic idea of dense pixel techniques is to map each dimension value to a colored pixel and group the pixels belonging to each dimension into adjacent areas. See figure 4.

*4) Stacked Displays*

Stacked display techniques are tailored to present data partitioned in a hierarchical fashion. In the case of multidimensional data, the data dimensions to be used for partitioning the data and building the hierarchy have to be selected appropriately. See figure 5.

### B. INTERACTION AND DISTORTION TECHNIQUES

In addition to the visualization technique, for an effective data exploration, it is necessary to use some interaction and distortion techniques. Interaction techniques allow the data analyst to directly interact with the visualizations and dynamically change the visualizations according to the exploration objectives and they also make it possible to relate and combine multiple independent visualizations. Distortion techniques help in the data exploration process by providing means for focusing on details while preserving an overview of the data. The basic idea of distortion techniques is to show portions of the data with a high level of detail, while others are shown with a lower level of detail.

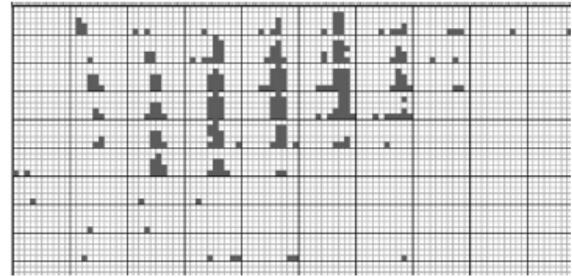

Figure 5. Dimensional Staking display (Courtesy IEEE)

*1) Dynamic Projections*

The basic idea of dynamic projections is to dynamically change the projections in order to explore a multidimensional data set. A classic example is the GrandTour system [24], which tries to show all interesting two-dimensional projections of a multidimensional data set as a series of scatter plots.

*2) Interactive Filtering*

In exploring large data sets, it is important to interactively partition the data set into segments and focus on interesting subsets. This can be done by a direct selection of the desired subset (browsing) or by a specification of properties of the desired subset (querying). Browsing is very difficult for very large data sets and querying often does not produce the desired results. Therefore, a number of interaction techniques have been developed to improve interactive filtering in data exploration. Examples are Magic Lenses [26], InfoCrystal [27] etc.

*3) Interactive Zooming*

In dealing with large amounts of data, it is important to present the data in a highly compressed form to provide an overview of the data, but, at the same time, allow a variable display of the data on different resolutions. Zooming not only means to display the data objects larger, but also means that the data representation automatically changes to present more details on higher zoom levels. The objects may, for example, be represented as single pixels on a low zoom level, as icons on an intermediate zoom level, and as labeled objects on a high resolution. Examples are: TableLens approach [28], PAD++ [29] etc.

*4) Interactive Distortion*

Interactive distortion techniques support the data exploration process by preserving an overview of the data during drill-down operations. The basic idea is to show portions of the data with a high level of detail while others are shown with a lower level of detail. Popular distortion techniques are hyperbolic and spherical distortions, which are often used on hierarchies or graphs, but may be also applied to any other visualization technique. An example of spherical distortions is provided in the Scalable Framework paper (see Fig. 5 in [23]). Other examples are Bifocal Displays [30], Graphical Fisheye Views [31] etc.

*5) Interactive Linking and Brushing*

The idea of linking and brushing is to combine different visualization methods to overcome the shortcomings of single techniques. It can be applied to visualizations generated by all visualization techniques described above. As a result, the brushed points are highlighted in all visualizations, making it possible to detect dependencies and correlations. Interactive changes made in visualization are automatically reflected in the other visualization. Typical examples of visualization techniques which are combined by linking and brushing are multiple scatterplots, bar charts, parallel coordinates, pixel displays, and maps. Most interactive data exploration systems allow some form of linking and brushing. Examples are Polaris [22], XGobi [25] and DataDesk [32].

Experiments with the real world problems showed that the visualization can significantly increase the quality and usefulness of web log mining results. However, how decision makers perceive and interact with a visual representation can strongly influence their understanding of the data as well as the usefulness of the visual presentation. In section III we try to explore the human aspects in visualization. In section IV we discuss some research examples.

## II. RELATED WORK

Most common technique of visualization is Graph drawing and it has been subject of research since decades [5, 9]. Graphs are a natural means to model the structure of the web, as the pages are represented by nodes and the links represented by edges. Many graph algorithms are used, in original or adapted form, to calculate and express properties of web sites and individual pages [4, 7, 8]. Although to a lesser extent, graph theoretic methods have also been applied to the user navigation paths through web sites [10]. WebQuilt is a logging and visualization system [11] which is interactive in the sense that it provides semantic zooming and filtering, given a storyboard. Webviz [2], VISVIP [3], VisualInsights [12] are some other visualization tools. So many commercial visualization tools for representing association rules have also been developed. Some of them are MineSet [14] and QUEST [13]. Becker [15, 16] describes a series of elegant visualization techniques designed to support data mining of business databases. Westphal et al. [17] give an excellent introduction of visualization techniques provided by current data mining tools. Cockburn and McKenzie [6] mention various issues related to graphical representations of web browsers' revisitation tools.

How a viewer perceives an item in a visualization display depends on many factors, including lighting conditions, visual acuity, surrounding items, color scales, culture, and previous experience [18]. There are so many technical challenges in developing a good visualization tool one of the big challenges is ***User acceptability.*** Much novel visualization techniques have been presented, yet their widespread deployment has not taken place, because of user acceptability due to lack of visual analytics approach. Many researchers have started their work in this direction. An example is the IBM Remail project [20] which tries to enhance human capabilities to cope with email overload. Concepts such as "Thread Arcs", "Correspondents Map", and "Message Map" support the user in efficiently analyzing his personal email communication. MIT's project Oxygen [19] even goes one step further, by addressing the challenges of new systems to be pervasive, embedded, nomadic, adaptable, powerful, intentional and eternal. Users are an integral part of the visualization process, especially when the visualization tool is interactive. Rheingans suggests that interaction should not be simply a "means to the end of finding a good representation" [21]. Interaction itself can be valuable since exploration may reveal insight that a set of fixed images cannot. Human factors-based design involves designing artifacts to be usable and useful for the people who are intended to benefit from them. Unfortunately, this principle is sometimes neglected in visualization systems.

## III. HUMAN FACTORS

How people perceive and interact with a visualization tool can strongly influence their understanding of the data as well as the system's usefulness. Human factors (e.g. interaction, cognition, perception, collaboration, presentation, and dissemination) play a key role in the communication between human and computer therefore contribute significantly to the visualization process and should play an important role in the design and evaluation of visualization tools. Several research initiatives have begun to explore human factors in visualization.

*A. Testing of Human Factors*

There are so many Human Computer Interaction interfaces available. Each interface is tested for its functionality (usability study) and ease of interaction (user studies).

*1) Ease of interaction*

To test ease of interaction we consider only real users and obtain both qualitative and quantitative data. Quantitative data typically measures task performance e.g. time to complete a specific task or accuracy e.g. number of mistakes. User ratings on questions such as task difficulty or preference also provide quantitative data. Qualitative data may be obtained through questionnaires, interviews, or observation of subjects using the system.

Walenstein [45] describes several challenges with formal user studies. According to him the main problem in the user studies is that we studies so many users but the true facts about the ease and benefits can be told only by the experts who can be difficult to find or may not have time to participate in lengthy studies. Another problem is that missing or inappropriate features in the test tool or problems in the interface can easily dominate the results and hide benefits of the ideas we really want to test. Thus, it seems that user studies can only be useful with an extremely polished tool so that huge amounts of time must be invested to test simple ideas that may not turn out to be useful. One solution to this problem is to have user studies focus on design ideas rather than complete visualization tools and to test specific hypotheses [45]. Our test should attempt to validate 1) whether the idea is effective and 2) why it is or is not effective. Of course, this may not be as easy as it sounds.

*2) Usability Study*

Additional evaluation methods established in Human Computer Interaction include cognitive walk-throughs (where an expert "walks through" a specific task using a prototype system, thinking carefully about potential problems that could occur at each step) and heuristic evaluations (where an expert evaluates an interface with respect to several predefined heuristics) [42]. Similarly, Blackwell et al. describe cognitive dimensions, a set of heuristics for evaluating cognitive aspects of a system [34], and Baldonado et al. designed a set of heuristics specific to multiple view visualizations [33]. These usability inspection methods avoid many of the problems with user studies and may be beneficial for evaluating visualizations. However, because these techniques are (for the most part) designed for user interface testing, it is not clear how well they will evaluate visualization ideas. For example, many visualization tasks are ill-defined. Walking through a complex cognitive task is very different from walking through a well-defined interface manipulation task. Furthermore, by leaving end users out of the evaluation process, usability inspection methods limit our ability to find unexpected errors.

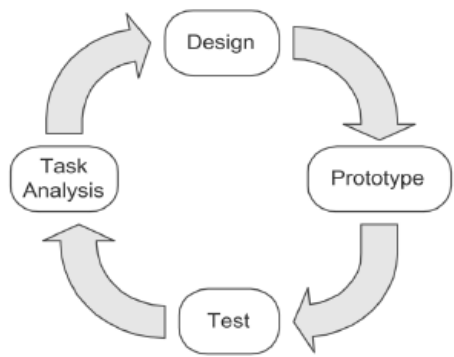

Figure 6. Visualization Design cycle

*B.   User-Centered Design*

User-centered design is an iterative process involving task analysis, design, prototype implementation, and testing, as illustrated in Fig. 6. Users are involved as much as possible at each design phase. Development may start at any position in the cycle, but would typically start with an analysis of the tasks the system should perform or testing of an existing system to determine its faults and limitations. User-centered design is more a philosophy than a specific method. Although it is generally accepted in human computer interaction, we believe this approach is not currently well-known in visualization and could support better visualization design. Various aspects of human factors-based design have been incorporated into visualization research and development. We provide examples of these contributions throughout the next section.

IV.   RESEARCH EXAMPLES

Adoption of human factors methodology and stringent evaluation techniques by the visualization community is in its infancy. A number of research groups have begun to consider these ideas and incorporate them into the design process to greater or lesser extents. This section will summarize these human factors contributions.

*A.   Improving Perception in Visualization Systems*

Several papers have looked at how our knowledge of perception can be used to improve visualization designs. For example, depth of focus is the range of distances in which objects appear sharp for a particular position of the eye's lens. Objects outside this range will appear blurry. Focusing effects can be used to highlight information by blurring everything except the highlighted objects [40]. For example, in computer games like road race the objects that are to be shown far are blurred giving the impact that object are far away and as the bike moves forward the blurring effect is reduced gradually giving impact of bike reaching near to the objects. Similarly in GIS application, all routes between two cities except for the shortest one could be blurred to highlight the best route. Here, the goal of blurring is to highlight information, not to focus on objects in the center of a user's field of view. Hence, the blurred objects are not necessarily at similar depths, a difference from traditional "depth of focus" effects. Figure 7 and 8, showing how perception can be improved by blurring.

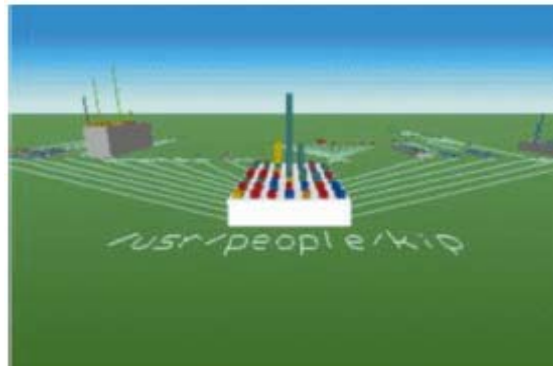

Figure 7.   Improving perception by blurring long distance objects

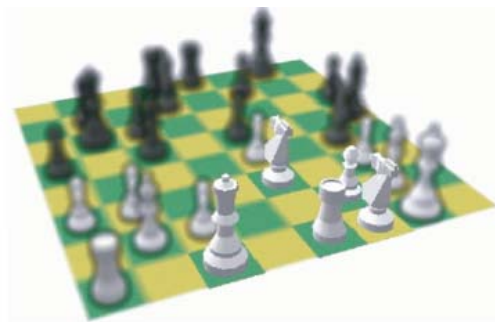

Figure 8.   Improving perception by blurring

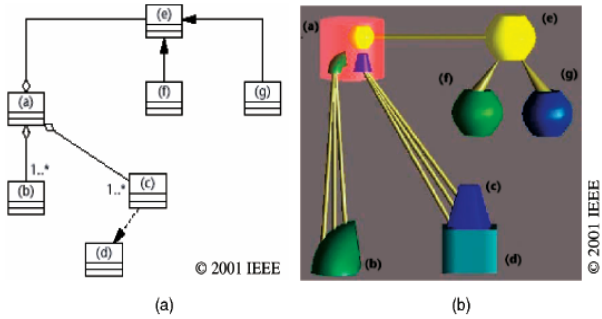

Figure 9. Perceptual Model

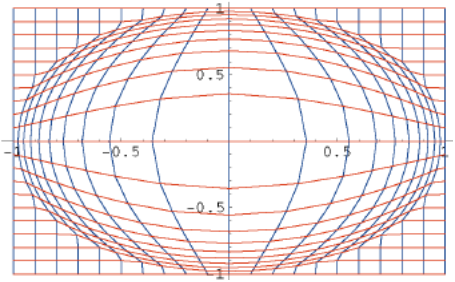

Figure 10. Fisheye distortion

*B. Interaction Metaphors*

Interacting with 3D visualizations can be challenging because mapping movements of a 2D mouse to actions in 3D space is not straightforward. Research has shown that manipulating objects relative to each other is easier than using absolute coordinates [37]. In addition, interaction may be easier when the interface is directly related to the task through task-specific props. Examples of task-specific props for visualization are: a physical model head and clip plane that aid interaction with volumetric brain data [38] and the "Cubic Mouse," a 3D input device for volume data that allows users to navigate along major axes by moving three perpendicular rods in a physical box [36]. Development of task-specific input devices for other visualization applications (e.g., flow visualization) could make interaction easier and thereby enhance data analysis.

In addition to the interactive hardware some interactive programming/presentation effort should be done for such a task like manipulating windows and widgets, navigating around interfaces and managing data, these tasks are called maneuvering. For example, an analyst examining user access to a website may begin by examining several visual images. Generating these images may require manipulation of several windows and widgets within the visualization tool. If the analyst then decides to examine the data quantitatively, he or she may need to return to the original window to look up values and/or switch to a different computer program in order to perform a mathematical analysis or generate statistics. These maneuvering operations are time consuming and distract users from their ultimate goals; thus, some necessary tools for these tasks should be integrated with the visualization tool to minimizing unnecessary navigation.

*C. Perceptual Models for Computer Graphics*

Various mathematical models of visual perception are available today. Typical models approximate contrast sensitivity, amplitude nonlinearity (sensitivity changes with varying light level), and masking effects of human vision. Two examples are the Daly Visual Differences Predictor [35] and the Sarnoff Visual Discrimination Model [41]. Variations on these models have been used for realistic image synthesis. Improving realism is not too much important in visualization because emphasis is not on representing the real world image but on representing data for the analysis purpose. Applications more relevant to visualization include increasing rendering speed (to enable interactive data exploration) and reducing image artifacts (to enhance perception and prevent incorrect interpretations of data). Reddy removed imperceptible details to reduce scene complexity and improve rendering speed [43].

*D. Transfer Functions*

In direct volume rendering, each voxel (sample in a 3D volume grid) is first classified as belonging to a particular category based on its intensity and/or spatial gradient value(s). Voxels are then assigned a color and transparency level based on this classification. The function that does this is called a transfer function. One example in Computed Tomography (CT) data would be to make skin semitransparent and bones opaque so the bones could be seen beneath the skin. In this case, transfer function design is quite easy since bones and skin have very different intensity values in CT data and can be easily distinguished. However, in general, finding good transfer functions is difficult and is therefore a major research area in volume visualization.

*E. Detail and Context Displays (Distortion)*

Resolution of the computer monitor is limited. Only a limited number of graphic items can be displayed at one time. Displaying more items often means displaying less detail about each item. If all items are displayed, few details can be read, but if only a few items are shown, we can lose track of their global location. Interactive distortion techniques support the data exploration process by preserving an overview of the data during drill-down operations. The basic idea is to show portions of the data with a high level of detail while others are shown with a lower level of detail.

*F. User and Computer Cooperation*

Computers can easily store and display data, but humans are better at interpreting data and making decisions. Although this idea is very useful, it is possible for computers to play a more active role in the visualization process than simply presenting data and providing an interface for data manipulation. As viewers look at images, they compare the image with their existing mental model of the data and presentation method and adjust either their mental model or their understanding of the image if the two conflict.

For complex data, constructing a mental model requires interaction and time since all the data cannot be seen in a single view. Allowing users to write down and manipulate their mental models, ideas, and insight (e.g., as mind maps) could reduce demands on human memory and help users identify new patterns or relationships.

## V. CONCLUSION AND FUTURE WORK

Scientists are utilizing visualization tools for doing data analysis in several disciplines. But the current visualization tools did not support "integration of insight," an important data analysis task involving taking notes, recording and organizing ideas and images, keeping track of the data analysis history, and sharing ideas with others. Overall, visualization systems could play several roles:

(a). Visually represent data to enhance data analysis,
(b). Visually display users' mental models, interpretations of the data, ideas, hypotheses, and insight,
(c). help users to improve their mental models by finding supporting and contradictory evidence for their hypotheses, and
(d). help users organize and share ideas.

Current research in visualization is almost exclusively devoted to the first objective. Research into the others has not been greatly explored and could make a valuable addition to data analysis tools. In the above study we identify several specific directions for future work. These are

- How to integrate human factors (perception and cognition theories) in the visualization techniques?
- Developing and evaluating task-specific input devices to aid interaction,
- Developing tools that provide cognitive support for insight and organization of ideas.

ACKNOWLEDGMENT

Author is grateful to the technical reviewers for the comments, which improved the clarity and presentation of the paper.

## AUTHORS PROFILE

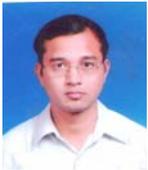
**Ratnesh Kumar Jain** is Ph. D. student at Dr. H. S. Gour Central University (formerly, Sagar University) Sagar, M P, India. He completed his bachelor's degree in Science (B. Sc.) with Electronics as special subject in 1998 and master's degree in computer applications (M.C.A.) in 2001 from Dr. H. S. Gour University, Sagar, MP, India. His field of study is Operating System, Data Structures, Web mining, and Information retrieval. He has published more than 5 research papers and has authored a book.

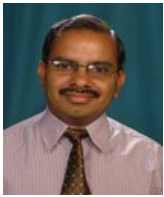
**Suresh Jain** completed his bachelor's degree in civil engineering from Maulana Azad National Institute of Technology (MANIT) (formerly, Maulana Azad College of Technology) Bhopal, M.P., India in 1986. He completed his master's degree in computer engineering from S.G. Institute of Technology and Science, Indore in 1988, and doctoral studies (Ph.D. in computer science) from Devi Ahilya University, Indore. He is professor of Computer Engineering in Institute of Engineering & Technology (IET), Devi Ahilya University, Indore. He has experience of over 21 years in the field of academics and research. His field of study is grammatical inference, machine learning, web mining, and information retrieval. He has published more than 25 research papers and has authored a book.

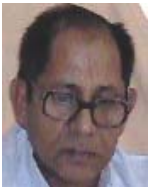
**R. S. Kasana** completed his bacholar's degree in 1969 from Meerut University, Meerut, UP, India. He completed his master's degree in Science (M.Sc.-Physics) and master's degree in technology (M. Tech.-Applied Optics) from I.I.T. New Delhi, India. He completed his doctoral and post doctoral studies from Ujjain University in 1976 in Physics and from P. T. B. Braunschweig and Berlin, Germany & R.D. Univ. Jabalpur correspondingly. He is a senior Professor and HoD of Computer Science and Applications Department of Dr. H. S. Gour University, Sagar, M P, India. During his tenure he has worked as vice chancellor, Dean of Science Faculty, Chairman Board of studies. He has more than 34 years of experience in the field of academics and research. Twelve Ph. D. has awarded under his supervision and more than 110 research articles/papers has published.